\documentclass[12pt]{article} 

\usepackage[symbol]{footmisc}
\renewcommand{\thefootnote}{\fnsymbol{footnote}}

\usepackage{comment}
\usepackage{latexsym}
\usepackage{amssymb,amsfonts}

\usepackage[utf8]{inputenc}
\usepackage[T1]{fontenc} 
\usepackage[english]{babel}           

\usepackage[pdftex]{graphicx}
\usepackage{epsfig}
\usepackage{graphicx}
\usepackage{comment}
\usepackage{latexsym}
\usepackage{hyperref}
\usepackage{amsmath}
\usepackage{mathrsfs}
\usepackage{calrsfs} 
\DeclareMathAlphabet{\pazocal}{OMS}{zplm}{m}{n} 
\usepackage[usenames, dvipsnames]{color}
\usepackage{amsbsy}
\usepackage{amssymb}
\usepackage{amsthm}
\usepackage{amsfonts}
\usepackage{cite}
\usepackage{enumitem}
\usepackage{xcolor}
\usepackage{color}
\usepackage{mathtools}
\usepackage{caption} 
\usepackage{subcaption} 
\usepackage{esvect}
\usepackage{cancel}

\usepackage{diagbox}

\usepackage{tikz}

\usepackage{tcolorbox}

\newcommand{\be}[0]{\begin{equation}}
\newcommand{\ee}[0]{\end{equation}}

\renewcommand{\thefootnote}{\fnsymbol{footnote}}

\newcommand{\Z}{\mathbb{Z}}
\renewcommand{\natural}{\mathbb{N}}

\renewcommand{\O}{{\pazocal O}}

\newcommand{\tr}{\textrm{tr}\,}

\renewcommand{\d}{\text{d}}
\newcommand{\e}{e}

\newcommand{\where}{\mbox{where}}

\renewcommand{\and}{\mbox{and}}

\newcommand{\espDD}{\phantom{\!\!\underset{\displaystyle |}{|}}}

\newcommand{\bm}{\boldmath} 

\newcommand{\N}{{\pazocal N}}
\renewcommand{\L}{{\pazocal L}}
\newcommand{\A}{{\pazocal A}}

\newcommand{\C}{{\pazocal C}}
\newcommand{\cR}{{\pazocal R}}

\newcommand{\m}{{\vec m}}
\newcommand{\vell}{{\vec \ell}}

\catcode`\@=11
\def\marginnote#1{}
\newcount\hour
\newcount\minute
\newtoks\amorpm
\hour=\time\divide\hour by60 \minute=\time{\multiply\hour by60
\global\advance\minute by-\hour}
\edef\standardtime{{\ifnum\hour<12 \global\amorpm={am}%
        \else\global\amorpm={pm}\advance\hour by-12 \fi
        \ifnum\hour=0 \hour=12 \fi
        \number\hour:\ifnum\minute<10 0\fi\number\minute\the\amorpm}}
\edef\militarytime{\number\hour:\ifnum\minute<10 0\fi\number\minute}
\def\draftlabel#1{{\@bsphack\if@filesw {\let\thepage\relax
   \xdef\@gtempa{\write\@auxout{\string
      \newlabel{#1}{{\@currentlabel}{\thepage}}}}}\@gtempa
   \if@nobreak \ifvmode\nobreak\fi\fi\fi\@esphack}
        \gdef\@eqnlabel{#1}}
\def\@eqnlabel{}
\def\@vacuum{}
\def\draftmarginnote#1{\marginpar{\raggedright\scriptsize\tt#1}}
\def\draft{\oddsidemargin -.2truein
        \def\@oddfoot{\sl preliminary draft \hfil
        \rm\thepage\hfil\sl\today\quad\militarytime}
        \let\@evenfoot\@oddfoot \overfullrule 3pt
        \let\label=\draftlabel
        \let\marginnote=\draftmarginnote
   \def\@eqnnum{(\theequation)\rlap{\kern\marginparsep\tt\@eqnlabel}%
\global\let\@eqnlabel\@vacuum}  }
\def\thebibliography#1{
\vskip 0.5cm \centerline{\bf \Large References}
\list{
[\arabic{enumi}]}{\settowidth\labelwidth{[#1]}
\leftmargin\labelwidth
\advance\leftmargin\labelsep
\usecounter{enumi}}
\def\newblock{\hskip .11em plus .33em minus .07em}
\sloppy\clubpenalty4000\widowpenalty4000
\sfcode`\.=1000\relax}

\renewcommand{\theequation}{\arabic{section}.\arabic{equation}}
\renewcommand{\section}{\setcounter{equation}{0}\@startsection
{section}{1}{0mm}{-\baselineskip}{0.5\baselineskip} {\normalfont\Large\bfseries}}
\renewcommand{\subsection}{\@startsection
{subsection}{2}{0mm}{-\baselineskip}{0.5\baselineskip} {\normalfont\large\bfseries}}
\renewcommand{\subsubsection}{\@startsection
{subsubsection}{3}{0mm}{-\baselineskip}{0.5\baselineskip}
{\normalfont\normalsize\slshape}}

\begin{document}


\begin{titlepage}
\begin{flushright}
CPHT-PC103.122019, December   2019
\vspace{1cm}
\end{flushright}

\begin{centering}
{\bm\bf \Large Induced cosmological and Newton constants \\ \vspace{.2cm} from towers of states  \footnote{Based on a talk given at the Corfu Summer Institute 2019 ``School and Workshops on Elementary Particle Physics and Gravity'' (CORFU2019), 31 August -- 25 September 2019, Corfu, Greece.}
 }

\vspace{5mm}

 {\bf Herv\'e Partouche and Balthazar de Vaulchier}

 \vspace{3mm}

 {CPHT, CNRS, Ecole polytechnique, IP Paris, \\F-91128 Palaiseau, France\\
{\em herve.partouche@polytechnique.edu, \\ balthazar.devaulchier@polytechnique.edu}}

\end{centering}
\vspace{0.7cm}
$~$\\
\centerline{\bf\Large Abstract}\\

\begin{quote}

We consider quadratic gravity with only Ricci squared and Weyl squared kinetic terms at tree-level, coupled to specific numbers of infinite towers of free scalars, Weyl fermions and gauge bosons in four dimensions. We show that predictable cosmological and Newton constants are induced at 1-loop. This result is cross-checked using both the heat kernel method and Adler's approach.

\end{quote}

\end{titlepage}
\newpage
\setcounter{footnote}{0}
\renewcommand{\thefootnote}{\arabic{footnote}}
 \setlength{\baselineskip}{.7cm} \setlength{\parskip}{.2cm}

\setcounter{section}{0}


\section{Introduction}

An appealing idea is that Einstein's theory may not be fundamental, but induced by quantum fluctuations of the matter content of the universe~\cite{zah,Adler1,Adler2,Adler3,Adler4,Adler5,Visser}. If this is the case, the dynamics of the spacetime metric  $g_{\mu\nu}$ is seen as an approximation of the dynamics of the underlying degrees of freedom. The fundamental theory is assumed to be free of UV divergences,  whereas the non-renormalizability of gravity is just an artefact of the effective theory.

In some approaches of induced gravity,  the metric is considered to be purely classical, a background field that  acquires a kinetic term only through loop corrections of the matter fields~\cite{zah,Adler1,Adler2,Adler3,Adler4,Adler5}. However, this assumption leads to several drawbacks, one of which being the absence of natural reason to impose the extremization of the effective action with respect ot $g_{\mu\nu}$: The Einstein equations should be imposed by hand. Another reason is that gravitational higher derivative terms are also generated, and it appears to be impossible to obtain finite results for the induced $\cR^2$- and $\C^2$-terms, where $\cR$ is the Ricci scalar and $\C_{\mu\nu\lambda\omega}$ is the Weyl tensor. These four-derivative terms need to be renormalized, which requires the existence of counterterms of the same form at tree-level.

In the present note, we consider infinite towers of matter fields (scalars, spinors and vectors) that interact only with a four-dimensional spacetime metric, whose classical kinetic terms contain exclusively four derivatives of the form $\cR^2$ and $\C^2$. The full theory is renormalizable~\cite{Stelle1,Stelle2,AG,Salvio}. If the matter fields can formally be seen as Kaluza-Klein (KK) states arising from a $(4+n)$-dimensional theory, the metric has no KK excitations and only makes sense from the four-dimensional point of view. We find that a cosmological constant $\Lambda_{\rm ind}$ and  a Newton constant $G_{\rm ind}$ are induced by the radiative corrections of the matter fields. They are finite and calculable from data of the underlying fundamental theory, provided the numbers of towers satisfy some specific rule. $\Lambda_{\rm ind}$ and $G_{\rm ind}$ remain predictable when quantum corrections of the gravitational degrees of freedom are taken into account at the 1-loop level. 

These results are derived in Sect.~\ref{HK} using heat kernel methods~\cite{zah,dewitt,BD,Avramidi2,Avramidi3}, and they are in perfect agreement with Adler's approach to induced gravity~\cite{Adler1,Adler2,Adler3,Adler4,Adler5} we consider in Sect.~\ref{AIE}. Further details beyond the results presented here can be found in Ref.~\cite{KPV}


\section{Induced gravitational terms from the heat kernel method}
\label{HK}
In this section, we first derive the effective gravitational couplings induced by the radiative corrections associated with infinite numbers of real scalars, Weyl fermions and vectors fields. Then, we will  justify why the form of the induced effective action remains valid once quantum correction in the gravitational sector are taken into account at 1-loop. The ``matter'' fields are organized in towers of KK modes arising from a $(4+n)$-dimensional spacetime of the form $M^4\times S^1\times \cdots \times S^1$, where the $S^1$ circles have radii $R_i$, $i\in\{4,\dots 3+n\}$, and $M^4$ is a four-dimensional spacetime of metric $g_{\mu\nu}$.\footnote{Our convention for its signature is $(-,+,+,+)$.}   The towers of states are labelled by an index $u$, such that 
\be
\begin{aligned}
&u\in \Phi\equiv \{1,\dots,N_\phi\} && \mbox{for real scalar fields}\, ,\\
& u\in \Psi\equiv\{ N_\phi+1,...,N_\phi+N_\psi\}&&\mbox{for  Weyl fermions}\, ,\\
& u\in\A\equiv\{N_\phi+N_\psi+1,...,N_\phi+N_\psi+N_A\}&&\mbox{for vector bosons}\, .
\end{aligned}
\ee 
For every tower $u$, the KK modes are labelled by  $\m\equiv (m_4,...,m_{3+n})\in\Z^n$ and their  masses squared are  given by 
\be
\label{mass}
M_{u\m}=\sum_{i=4}^{3+n}\Big( \frac{m_i+Q_{ui}}{R_i}\Big)^2\, .
\ee
In this expression, $\vec Q_u$ is a real $n$-vector, whose components shift the KK momenta. From a $(4+n)$-perspective, $\vec Q_u$ is a charge vector under the Abelian isometries of the $n$ circles.\footnote{In that case, the towers should be combined in pairs in order to be complexified.} Even if this is not compulsory, we will assume that $\vec Q_u\notin \Z^n$ for all $u$, so that no massless state belongs to the KK spectrum.  

The classical action of the full theory is
\be
S_{\text{tree}} = S_{\rm g} + \sum_u S_u(\vec{Q}_u)\, ,
\label{tac}
\ee
where $S_{\rm g}$ is a purely gravitational action and $S_u(\vec{Q}_u)$ depends on the nature of the tower $u$. For $u\in\Phi$, we have 
\be
S_u(\vec Q_u)=-\int \d^4x\sqrt{-g}\,  \frac{1}{2}\sum_\m \Big[g^{\mu\nu}\partial_\mu\phi_{u\m}\partial_\nu\phi_{u\vec m} + M_{u\m}^2\phi_{u\m}^2 \Big],
\label{scalar}
\ee
where $\phi_{u\m}$ is a real scalar. For $u\in\Psi$, the action is 
\be
\label{fermion}
S_u(\vec Q_u)= \int\d^4x \sqrt{-g}\,  \frac{1}{2}\sum_\m \Big[i\nabla_\mu\bar \psi_{u\m} \bar \sigma^\mu \psi_{u\m}-i\bar \psi_{u\m} \bar \sigma^\mu\nabla_\mu\psi_{u\m}-M_{u\m} \big(\psi_{u\m}\psi_{u\m}+\bar \psi_{u\m}\bar \psi_{u\m}\big)\Big] ,
\ee
where the conventions for Weyl spinors we use are those of Ref.~\cite{wb}. Finally, for $u\in \A$, the action of the tower of vector bosons $A_{u\m}$ is 
\be
\label{vector}
S_u(\vec Q_u)= -\int\d^4x \sqrt{-g}\, \sum_\m \Big[ \frac{1}{4}\, g^{\mu\rho}g^{\nu\sigma}F_{u\m\mu\nu}F_{u\m\rho\sigma}+{1\over 2}\, M_{u\m}^2\, g^{\mu\nu}A_{u\m\mu}A_{u\m\nu}\Big],
\ee
where $F_{u\m\mu\nu}\equiv\partial_\mu A_{u\m\nu}-\partial_\nu A_{u\m\mu}$ is the field strength. In unitary gauge, the ghosts are infinitely massive and do not need to be considered when computing radiative corrections. 

Following Ref.~\cite{BD}, the 1-PI effective action in the semiclassical limit for $g_{\mu\nu}$ can be written as 
\be
S_{\text{eff}} = S_{\rm tree} + W\, ,
\ee
where $W$ arises from the radiative corrections of the matter fields,
\be
W = \frac{i}{2}\sum_u(-1)^F\sum_\m \ln \det \Big[(-D_u^2+M_{u\m}^2-i\varepsilon)\sigma^2\Big].
\ee
In our notations, $F=0$ for $u\in \Phi\cup \A$ and $F=1$ for $u\in \Psi$, while  $D_u^2$ is the kinetic operator of the KK modes of the tower $u$ and $\sigma$ is an arbitrary length introduced for dimensional purpose.
Using the identity $\ln\det=\tr\ln$ and the relation
\be
\ln(K\sigma^2) = -\int_{\rho^2}^{+\infty}\frac{\d s}{s}\e^{-isK} - \ln\Big[i\e^\gamma\Big(\frac{\rho}{\sigma}\Big)^2\Big] + \O(\rho^2 K) \, ,
\ee
which is valid for any complex number $K$ having a small negative imaginary part $-i\varepsilon$, we obtain
\be
\begin{aligned}
W&=-{i\over 2}\sum_u (-1)^F\sum_\m \tr \int_{\rho^2}^{+\infty}{\d s\over s}\, \e^{-is(-D_u^2+M_{u\m}^2-i\varepsilon)}+W_1\, ,  \\
\where \quad  W_1&=\phantom{-}{i\over 2}\sum_u (-1)^F\sum_\m \tr\Big( \ln\Big[i\e^\gamma\Big(\frac{\rho}{\sigma}\Big)^2\Big] + \O(\rho^2(-D_u^2+M_{u\m}^2-i\varepsilon))\Big) . 
\end{aligned}
\ee
Note that in $W_1$, the term proportional to $\ln(i\e^\gamma (\rho/\sigma)^2)$ cancels out when the number of bosonic and fermionc towers are equal, $N_\phi+N_A=N_\psi$. Indeed, we will impose later on this condition in order to eliminate all divergencies occurring in the cosmological constant. Anticipating this fact, and noticing that the contribution $\O(\rho^2)$ vanishes in the UV limit $\rho\to 0$, we  discard from now on the contribution $W_1$. The relevant term in the expression of $W$ can be computed thanks to the heat kernel expansion\cite{dewitt,BD,Avramidi2,Avramidi3}
\be
-{i\over 2}\, \tr \e^{-is(-D_u^2+M_{u\m}^2-i\varepsilon)} =-{1\over 2} \int \d^4x \sqrt{-g} \;  {\e^{-is(M_{u\m}^2-i\varepsilon)}\over (4\pi s)^2} \, \sum_{\kappa=0}^{+\infty}\,(is)^\kappa a_{u\kappa}\, , 
\ee
where the $a_{u\kappa}$ are functions of the local spacetime geometry. Dimensional analysis implies that they contain $2\kappa$ derivatives. For the lower values of $\kappa$, we have
\be
\begin{aligned}
a_{u0}&=k_{u\Lambda}\, , \\
a_{u1}&=k_{u\cR}\, \cR\, , \\
\sqrt{-g}\, a_{u2}&=\sqrt{-g}\,\Big[k_{u\cR^2}\,\cR^2+ k_{u\C^2}\, \C^2 \Big]\!+\mbox{total derivative}\, ,
\end{aligned}
\label{fff}
\ee
where $\cR$ is the Ricci scalar and $\C^2\equiv \C_{\mu\nu\lambda\omega} \C^{\mu\nu\lambda\omega}$, with $\C_{\mu\nu\lambda\omega}$ the Weyl tensor. Moreover, the coefficients appearing in the r.h.s., which depend on the nature of the KK modes, are given in  Table~\ref{tabcoeff}~\cite{dewitt,BD,Duff1,Duff2}. 
\begin{table}
\begin{centering}
\begin{tabular}{l|cccc}
 & $k_{u\Lambda}$ & $k_{u\cR}$ & $k_{u\cR^2}$ & $k_{u\C^2}$ \\
\hline
tower $u$ of real scalars & $1$ & ${1\over 6}$ & ${1\over 72}$ & ${1\over 120} $\\
tower $u$ of  Weyl fermions & $2$ & $-{1\over 6}$ & $0$ & $-{1\over 40}  $\\
tower $u$ of  massive vector bosons & $3$ & $-{1\over 2}$ & ${1\over 72}$ & $ {13\over 120}$\\
\hline
\end{tabular}
\par\end{centering}
\caption{\footnotesize Coefficients $k_{u\Lambda},\, \, k_{u\cR},\, \, k_{u\cR^2}$ and $k_{u\C^2}$ appearing in Eq.~(\ref{fff}) for each mode of the KK tower $u$. }
\label{tabcoeff}
\end{table}
Finally, the total derivative, which contains in particular a contribution proportional to the Gauss--Bonnet density, will not be considered in the following because it only contributes as a constant to the action. Thus, $W$ can be decomposed as
\be
\begin{aligned}
&W=\int \d^4x \sqrt{-g} \,\sum_{\kappa=0}^{+\infty}\L_\kappa\, , \\
\where\quad &\L_\kappa(x)={1\over 32 \pi^2} \sum_u (-1)^{F} a_{u\kappa}(x)\sum_\m \int_{\rho^2}^{+\infty}i\d s \, f_{u\kappa \m}(is)\, , \\
& f_{u\kappa \m}(is)=(is)^{\kappa-3}\, \e^{-is(M_{u\m}^2-i\varepsilon)}  \, .
\end{aligned}
\label{Lags}
\ee

In order to compute radiative corrections of an infinite number of states, we make a choice of prescription encountered in string theory. Namely, we first sum over the spectrum and  then compute the Schwinger integral. The reason for this is that the sum over the spectrum may have a smoother behavior than the contribution of each individual state. Applying this rule to the expressions of the induced Langrangian densities $\L_\kappa$, it is then  legitimate to apply a Wick rotation, a change of variable $t=is$, and to take the limit $\rho\to 0$ to write 
\be
\L_\kappa(x)={1\over 32 \pi^2} \int_{0}^{+\infty}\d t \sum_u (-1)^{F} a_{u\kappa}(x) \sum_\m f_{u\kappa \m}(t)\, ,
\ee
provided the integrand has no pole~\cite{KPV}. To see when this is the case, let us consider three cases:

\begin{itemize}
\item For $\kappa\geq3$, the function $f_{u\kappa\m}$ has no pole and the result is
\be
\forall \kappa\ge 3:\quad \L_\kappa = {\Gamma(\kappa-2)\over 32\pi^2}\sum_u (-1)^{F} \sum_\m {1\over M_{u\m}^{2(\kappa-2)}}\, a_{u\kappa}\, .
\ee

\item For $\kappa=0,1$, a Poisson summation over $\m$ leads to 
\be
\sum_\m f_{u\kappa \m}(t)={\pi^{n\over 2} \over t^{3+{n\over 2}-\kappa}}\Big(\prod_i R_i\Big)\sum_{\vell\in\Z^n} \e^{-{\pi^2\over t} \sum_j(\ell_jR_j)^2}\e^{2i\pi \vec Q_u\cdot\vell}\, ,
\label{po}
\ee
which has a pole only for  $\vell=0$. However, the latter is independent of the charge $\vec Q_u$. Hence the sums over the towers $u$ of all unwanted contributions  $\vell=\vec 0$ in $\L_0$ and $\L_1$ are proportional to
\be
\sum_u(-1)^{F}k_{u\Lambda}  \qquad \mbox{and} \qquad \sum_u(-1)^{F}k_{u\cR} = 0 \, .
\ee
Using Table~\ref{tabcoeff}, these coefficients vanish when the numbers of towers satisfy the conditions 
\be
N_\phi -2N_\psi +3N_A = 0 \, , \qquad N_\phi+N_\psi-3 N_A=0\, , 
\ee
which admit the  solutions 
\be \label{spec}
(N_\phi,N_\psi,N_A) = (1,2,1)N \, , \quad N\in\natural.
\ee
In that case, the integrals over $t$ can be computed term by term using the change of variable $t=\pi^2\sum_j(\ell_jR_j)^2/l$, and the result is
\be\label{Gind}
\begin{aligned}
\L_0&\equiv-\frac{1}{8\pi}\frac{\Lambda_{\rm ind}}{G_{\rm ind}}\;= {\Gamma(2+{n\over 2})\over 32\pi^{6+{n\over 2}}}\Big(\prod_{i} R_i\Big)\sum_u (-1)^{F}k_{u\Lambda}\sum_{\vell\neq \vec 0}\frac{\e^{2i\pi \vec Q_u\cdot\vell}}{\big(\sum_j\ell_j^2R_j^2\big)^{2+{n\over2}}}\, ,\\
\L_1&\equiv \frac{1}{8\pi G_{\rm ind}}\,  {\cR\over 2}={\Gamma(1+{n\over 2})\over 16\pi^{4+{n\over 2}}}\Big(\prod_{i} R_i\Big)\sum_u (-1)^{F}k_{u\cR}\sum_{\vell\neq \vec 0}\frac{\e^{2i\pi \vec Q_u\cdot\vell}}{\big(\sum_j\ell_j^2R_j^2\big)^{1+{n\over2}}}\, {\cR\over 2}\, .
\end{aligned}
\ee

\item For $\kappa=2$, the Poisson summation leads again to a singular contribution $\vell=0$. Summing over $u$, the divergent coefficients of the $\cR^2$ and $\C^2$ terms are proportional to 
\be
\sum_u(-1)^{F}k_{u\cR^2}   \qquad \mbox{and}\qquad \sum_u(-1)^{F}k_{u\C^2}  \, .
\ee
Upon using Table~\ref{tabcoeff}, vanishing of these coefficients amounts to satisfying the constraints  
\be
N_\phi+N_A=0 \, , \qquad N_\phi +3N_\psi +13N_A = 0 \, ,
\ee
which however have no solution. Thus, the Lagrangian density $\L_2$ is UV divergent and must be renormalized thanks to tree-level $\cR^2$ and $\C^2$ counterterms. As a consequence, the purely gravitational classical action appearing in Eq.~(\ref{tac}) is defined  as~\cite{Stelle1,Stelle2,Salvio}
\be
S_{\rm g} = \int\d^4x\sqrt{-g}\left(\frac{\cR^2}{f_{0\rm B}^2} - \frac{\C^2}{f_{2\rm B}^2} \right)  ,
\ee
where $f_{0\rm B}^2$, $f_{2\rm B}^2$ are infinite bare couplings.  

\end{itemize}
Collecting all our results, the semiclassical 1-PI effective action reads
\be
\label{res}
S_{\text{eff}} = \int\d^4x\sqrt{-g} \left[{1\over 8\pi G_{\rm ind}}\Big({\cR\over 2}-\Lambda_{\rm ind}\Big)+\frac{\cR^2}{f_0^2} - \frac{\C^2}{f_2^2}+\sum_{\kappa=3}^{+\infty}\L_\kappa+\sum_{u=1}^{N_\phi+N_\psi+N_A}S_u(\vec Q_u)\right]\!,
\ee
where $f_0^2$ and $f_2^2$ are finite couplings to be determined by measurements.

Notice that at this stage, the gravitational fields are quantum degrees of freedom that have been treated semiclassically. However, the above expression remains valid once radiative corrections in the gravitational sector are taken into account at the 1-loop level. To understand why, notice that when no matter is coupled to pure quadratic gravity, namely the theory of gravity where the kinetic terms are of the form $\cR^2$ and $\C^2$ only, no cosmological and Newton constants are generated at the quantum level. This follows from the fact that scale invariance is a non-anomalous symmetry of this theory~\cite{Stelle1,Stelle2,Salvio,AG,CP,Salvio2}. In our case, though,  quadratic gravity is coupled to towers of matter fields. However, because \mbox{1-loop} Feynman diagrams arising from different particles are simply added linearly to each other, we conclude that Eq.~(\ref{res}) continues to be true at this order. Nevertheless, whether predictability of the induced cosmological and Newton constants  is preserved beyond \mbox{1-loop} involving gravitational degrees of freedom is a question that is beyond the scope of the present paper. 

Before concluding this section, we would like to present the simplest example where finiteness of the cosmological and Newton constants are achieved in our framework. This corresponds to the case  where $\N=1$ supersymmetry is spontaneously broken via implementation of a Scherk-Schwarz mechanism~\cite{SS1} along $n=1$ circle of radius $R_4$. If all bosons have a universal charge, we have 
\be
\mbox{for  $u\in\Phi\cup\A$ : }\; Q_u=Q\notin\Z\, ,\qquad \mbox{for $u\in\Psi$ : }\; Q_u=Q+{1\over 2}\, .
\ee
Using Eq.~(\ref{Gind}), we obtain
\be
\begin{aligned}
\frac{1}{8\pi}\frac{\Lambda_{\rm ind}}{G_{\rm ind}}= -N \, \frac{\Gamma\big({5/ 2}\big)}{4\pi^{{13\over2}}}\frac{1}{R_4^4}\sum_{\ell}\frac{\e^{2i\pi  Q(2\ell+1)}}{\big|2\ell+1\big|^{5}}\, ,\,\; \quad \frac{1}{8\pi G_{\text{ind}}} = N\,\frac{\Gamma(3/2)}{24\pi^{9\over2}}\frac{1}{R_4^2}\sum_\ell \frac{\e^{2i\pi Q(2\ell+1)}}{\big| 2\ell+1 \big|^3}\, .
\end{aligned}
\ee


\section{Induced Einstein action from Adler's approach}
\label{AIE}
An alternative approach to compute  the induced gravitational couplings was developed by Adler~\cite{Adler1,Adler2,Adler3,Adler4,Adler5}. Instead of using a heat kernel expansion in derivatives, one may define the variation of the metric near a flat background and Taylor expand the effective action arising by integrating out the massive fields. The two first coefficients of this expansion are the cosmological and Newton constants, which can be expressed as 
\be\label{adler_form}
\begin{aligned}
\frac{1}{8\pi}\frac{\Lambda_{\rm ind}}{G_{\rm ind}} =& -{1\over 4}\, \langle T(0)\rangle\, , \\
\frac{1}{8\pi G_{\rm ind}} =& -\frac{i}{48}\int\d^4x\; x^2\, \langle\widetilde{T}(x)\widetilde{T}(0)\rangle\, .
\end{aligned}
\ee
In these formulas, $T(x)$ is the trace of the stress-energy tensor of the massive fields to be integrated out, $\widetilde{T}(x)\equiv T(x)-\langle T(x)\rangle$, and the correlators of time-ordered operators are evaluated in  Minkowski spacetime. In the following, we derive in Adler's framework the Einstein action induced by $N_\phi$, $N_\psi$, $N_A$ KK towers of real scalar fields, Weyl fermions and vector bosons, and find perfect agreement with the results of the previous section. 

The traces in flat space of the stress-energy tensors of the towers of states can be derived from the actions~\eqref{scalar},~\eqref{fermion} and~\eqref{vector}: 
\begin{align}
\label{T}
T_\phi(x)=&\:-\sum_{u\in\Phi}\sum_\m \Big[\partial_\mu\phi_{u\m}\partial^\mu\phi_{u\m} + 2M_{u\m}^2\phi_{u\m}^2 \Big] , \nonumber\\
T_\psi(x) =&\quad\,\sum_{u\in\Psi}\sum_\m\left[\frac{3i}{2}\Big(\partial_\mu\bar\psi_{u\m}\bar\sigma^\mu\psi_{u\m} - \bar\psi_{u\m}\bar\sigma^\mu\partial_\mu\psi_{u\m}\Big) -2M_{u\m}\Big(\psi_{u\m} \psi_{u\m}+\bar\psi_{u\m}\bar\psi_{u\m}\Big)\right], \nonumber \\
T_A(x) =&\;-\sum_{u\in \A}\sum_\m\, M_{u\m}^2\, A_{u\m}^\mu A_{u\m\mu}\, .
\end{align}
The associated correlators involved in the cosmological term are
\be
\begin{aligned}
\langle T_\phi(0)\rangle =&\ \sum_{u\in\Phi}\sum_\m \big[ i\delta^{(4)}(0)-M_{u\m}^2\,\Delta_{u\m}(0) \big]  , \\
\langle T_\psi(0) \rangle =&\ \sum_{u\in\Psi}\sum_\m \big[ -6i\delta^{(4)}(0)+2M_{u\m}^2\,\Delta_{u\m}(0) \big]  , \\
\langle T_A(0)\rangle =&\ \sum_{u\in\pazocal{A}}\sum_\m\big[i\delta^{(4)}(0) -3M_{u\m}^2\, \Delta_{u\m} (0)\big] ,
\end{aligned}
\ee
where $\Delta_{u\m}(x-y)$ is the Feynman propagator of a real scalar field of mass squared $M_{u\m}^2$, while the $\delta^{(4)}$-functions at $x=0$ are  infinite constants that contain no physical information and are harmless. One way to see this is to consider an equivalent definition of the classical theory that includes  non-dynamical real scalar fields. The action associated with an infinite number of such fields $D_\m$ is 
\be\label{aux}
S_{\text{non-dyn}} = \int\d^4x\sqrt{-g}\, \sum_\m \frac{1}{2}\, D_\m^2 \, ,
\ee
which yields an extra contribution to the previous correlators equal to  $\sum_\m 2i\delta^{(4)}(0)$. Because we can choose the number of such towers so that all terms of this form cancel out~\cite{KPV}, we will discard all $\delta^{(4)}$-functions from now on.
In order to compute the pieces of the correlators that involve the propagators, we follow  a prescription inspired by what is commonly done in string theory: 
\begin{itemize}
\item We apply a Wick rotation and switch to first quantized formalism by introducing a Schwinger parameter $t$,
\be
\Delta_{u\m}(x)\equiv -i\int\frac{\d^4k}{(2\pi)^4}\frac{\e^{ik\cdot x}}{k^2+M_{u\m}^2-i\epsilon} = \int\frac{\d^4k_{\rm E}}{(2\pi)^4}\int_0^{+\infty}\d t\, \e^{-k^2_{\rm E}t+ik_{\rm E}\cdot x_{\rm E}}\e^{-M_{u\m}^2 t}\, .
\ee

\item We compute the Gaussian integral over the Euclidean momentum $k_{\rm E}$.

\item We perform a Poisson summation on the KK momentum $\m$.

\item Adding the contributions of all KK towers, the divergent terms (labelled by  $\vell=\vec 0$ as in the previous section) cancel out provided the condition~(\ref{spec}) is fulfilled. 
\item All other contributions $\ell\neq \vec 0$ can be integrated term by term over~$t$.
\end{itemize}
Applying this procedure, we find  that the induced cosmological constant obtained by adding the contributions of the integrated out scalars, fermions and gauge bosons is  
\be
\frac{1}{8\pi}\frac{\Lambda_{\rm ind}}{G_{\rm ind}}\Big|_\phi+\frac{1}{8\pi}\frac{\Lambda_{\rm ind}}{G_{\rm ind}}\Big|_\psi+\frac{1}{8\pi}\frac{\Lambda_{\rm ind}}{G_{\rm ind}}\Big|_A=\sum_{u\in\Phi}I'_0(\vec Q_u)-2\sum_{u\in\Psi}I'_0(\vec Q_u)+3\sum_{u\in\A}I'_0(\vec Q_u)\, , 
\ee
where we have defined 
\be
I'_0(\vec Q)=-\frac{\Gamma(2+{n\over2})}{32\pi^{6+{n\over2}}}\Big(\prod_i R_i\Big) \sum_{\vell\neq \vec 0} \frac{\e^{2i\pi\vec{Q}\cdot\vell}}{\big(\sum_j\ell_j^2R_j^2\big)^{2+{n\over2}}} \, .
\ee
This result is in agreement  with what we found in Eq.~\eqref{Gind}.

The derivation of the gravitational constant is more involved. The relevant correlators are expressed as
\begin{align}
\label{TT}
\langle\widetilde T_\phi(x)\widetilde T_\phi(0)\rangle =&\ 2\sum_{u\in\Phi}\sum_\m \Big[ \partial_\mu\partial_\nu\Delta_{u\m}(x)\,   \partial^\mu\partial^\nu\Delta_{u\m}(x)+ 4M_{u\m}^2\partial_\mu\Delta_{u\m}(x) \,\partial^\mu\Delta_{u\m}(x)\nonumber\\
&\ \qquad\qquad\,+ 4M_{u\m}^4\Delta_{u\m}(x)^2 \Big]  , \nonumber\espDD\\
\langle\widetilde T_\psi(x)\widetilde T_\psi(0)\rangle=&\ \sum_{u\in\Psi}\sum_\m\Big[ \!-9\big(\square\Delta_{u\m}(x)\big)^2 + 9\partial_\mu\Delta_{u\m}(x)\partial^\mu\square\Delta_{u\m}(x)  \nonumber\\
&\, + 39M_{u\m}^2\Delta_{u\m}(x)\square\Delta_{u\m}(x)- 7M_{u\m}^2\partial_\mu\Delta_{u\m}(x)\partial^\mu\Delta_{u\m}(x)- 32M_{u\m}^4\Delta_{u\m}(x)^2  \Big]  , \nonumber\espDD\\
\langle\widetilde{T}_A(x)\widetilde{T}_A(0)\rangle=&\ 2\sum_{u\in\pazocal{A}}\sum_\m\Big[  \partial_\mu\partial_\nu\Delta_{u\m}(x)\,   \partial^\mu\partial^\nu\Delta_{u\m}(x)  - 2M_{u\m}^2\Delta_{u\m}(x)\partial^2\Delta_{u\m}(x)\nonumber\\
&\ \qquad\quad \;\;\;\;+ 4M_{u\m}^4\Delta_{u\m}(x)^2 \Big] ,
\end{align}
which yield accordingly contributions given by 
\be
\begin{aligned}
\frac{1}{8\pi G_{\rm ind}}\Big|_\phi =&\ \sum_{u\in\Phi} \big[ I_1(\vec{Q_u})+I_2(\vec{Q}_u)+I_3(\vec{Q}_u) \big] , \\
\frac{1}{8\pi G_{\rm ind}}\Big|_\psi =&\ \sum_{u\in\Psi}\big[I_4({\vec Q}_u)  + I_5({\vec Q}_u) + I_6({\vec Q}_u) - \frac{7}{8}I_2({\vec Q}_u)- 4I_3({\vec Q}_u) \big] , \\
\frac{1}{8\pi G_{\rm ind}}\Big|_A =&\ \sum_{u\in\pazocal{A}}\big[I_1(\vec{Q}_u)-\frac{4}{39}I_6(\vec{Q}_u)+I_3(\vec{Q}_u) \big]  ,
\end{aligned}
\ee
where we have defined the integrals 
\begin{align}\label{integrals}
I_1({\vec Q}_u) =&\ \frac{i}{24\, (2\pi)^8}&&\!\!\!\!\!\!\int\d^4x\ x^2\sum_{\vec m}\int \d^4p\,\d^4k\ \frac{(p\cdot k)^2\; \e^{i(p+k)\cdot x}}{\big(p^2+M_{u\m}^2-i\varepsilon\big)\big(k^2+M_{u\m}^2-i\varepsilon\big)}\, , \nonumber \\
I_2({\vec Q}_u) =&\ -\frac{i}{6\, (2\pi)^8}&&\!\!\!\!\!\!\int\d^4x\ x^2\sum_{\vec m}M^2_{u\m}\int \d^4p\,\d^4k\ \frac{ p\cdot k\; \e^{i(p+k)\cdot x}}{\big(p^2+M_{u\m}^2-i\varepsilon\big)\big(k^2+M_{u\m}^2-i\varepsilon\big)}\, , \nonumber \\
I_3({\vec Q}_u) =&\ \frac{i}{6\, (2\pi)^8}&&\!\!\!\!\!\!\int\d^4x\ x^2\sum_{\vec m}M^4_{u\m}\int \d^4p\,\d^4k\ \frac{\e^{i(p+k)\cdot x}}{\big(p^2+M_{u\m}^2-i\varepsilon\big)\big(k^2+M_{u\m}^2-i\varepsilon\big)}\, , \nonumber\\
I_4({\vec Q}_u) =&\ -\frac{3i}{16\, (2\pi)^8}&&\!\!\!\!\!\!\int\d^4x\ x^2\sum_{\vec m}\int \d^4p\,\d^4k\ \frac{p^2\, k^2\; \e^{i(p+k)\cdot x}}{\big(p^2+M_{u\m}^2-i\varepsilon\big)\big(k^2+M_{u\m}^2-i\varepsilon\big)}\, ,  \\
I_5({\vec Q}_u) =&\ \frac{3i}{16\, (2\pi)^8}&&\!\!\!\!\!\!\int\d^4x\ x^2\sum_{\vec m}\int \d^4p\,\d^4k\ \frac{ p^2\, (p\cdot k)\; \e^{i(p+k)\cdot x}}{\big(p^2+M_{u\m}^2-i\varepsilon\big)\big(k^2+M_{u\m}^2-i\varepsilon\big)}\, , \nonumber \\
I_6({\vec Q}_u) =&\ -\frac{13i}{16\, (2\pi)^8}&&\!\!\!\!\!\!\int\d^4x\ x^2\sum_{\vec m}M^2_{u\m}\int \d^4p\,\d^4k\ \frac{p^2\; \e^{i(p+k)\cdot x}}{\big(p^2+M_{u\m}^2-i\varepsilon\big)\big(k^2+M_{u\m}^2-i\varepsilon\big)}\, . \nonumber 
\end{align}
These quantities can be computed by following the procedure described in the derivation of the cosmological constant, up to additional Gaussian  integratials over the Euclidean spacetime point~$x_{\rm E}$. The results are
\be
\begin{aligned}
I_1({\vec Q}) =&\; -\frac{\Gamma\big(1+{n\over 2}\big)}{32 \pi^{4+{n\over 2}}}\, \Big(\prod_i R_i\Big)\sum_{\vell}\frac{\e^{2i\pi {\vec Q}\cdot\vell}}{\big(\sum_j\ell_j^2R_j^2\big)^{1+{n\over2}}}\, , \\
I_2({\vec Q}) =&\ -\frac{4}{3}\, I_1({\vec Q})\, ,  \quad 
I_3({\vec Q}) =0\, , \quad
I_4(\vec{Q})=0 \, , \quad I_5(\vec{Q})=-\frac{3}{2}I_1(\vec{Q}) \, , \quad I_6(\vec{Q})=0 \, ,
\end{aligned}
\ee
where all ill-defined terms $\vell=\vec 0$ must cancel one another by the interplay between the towers of KK modes. Because this turns out to be the case  when Eq.~(\ref{spec}) is satisfied, the contributions induced by integrating out the scalars, fermions and gauge bosons to the Newton's constant add up to yield the finite result
\be
\frac{1}{8\pi G_{\rm ind}}\Big|_\phi+\frac{1}{8\pi G_{\rm ind}}\Big|_\psi+\frac{1}{8\pi G_{\rm ind}}\Big|_A=-{1\over 3}\sum_{u\in\Phi}I'_1(\vec Q_u)-{1\over 3}\sum_{u\in\Psi}I'_1(\vec Q_u)+\sum_{u\in\A}I'_1(\vec Q_u)\, , 
\ee
where $I'_1(\vec Q)$ is defined as $I_1(\vec Q)$, with the pathological term $\vell=\vec 0$ removed. As was the case for the cosmological constant, this result agrees with that found from the heat kernel method, Eq.~(\ref{Gind}).


\section*{Acknowledgements}

The work of H.P. is partially supported by the Royal-Society/CNRS International Cost Share Award IE160590.


\bibliographystyle{unsrt}


\end{document}